# Flat panel laser displays enabled by large-scale visible photonic integrated circuits

**Authors**

Zhujun Shi[†], Risheng Cheng[†,] Guohua Wei[†], Steven A. Hickman, Min Chul Shin, Peter Topalian, Lei Wang, Dusan Coso, Brian Le, Lizzy Lee, Sean Braxton, Alexander Koshelev, Maxwell F. Parsons, Rahul Agarwal, Barry Silverstein, Yun Wang, Giuseppe Calafiore[‡]

**Affiliations**

Reality Labs Research, Meta Platforms Inc., 10301 Willows Rd NE, Redmond, WA 98052

† Equally contributed to this work
‡ Corresponding author: giuseppe.calafiore@meta.com

**Abstract**

Laser-based displays are highly sought after for their superior brightness and color performance, especially in advanced applications like augmented reality (AR). However, their broader adoption has been hindered by bulky projector designs and complex optical module assemblies. Here, we introduce a new laser display architecture enabled by large-scale visible photonic integrated circuits (PICs) to address these challenges. Unlike previous projector-style laser displays, this architecture features an ultra-thin, flat-panel form factor, replacing bulky free-space illumination modules with a single, high-performance photonic chip. Centimeter-scale PIC devices, which integrate thousands of distinct optical components on-chip, are carefully tailored to achieve high display uniformity, contrast, and efficiency. We demonstrate a 2 mm-thick flat-panel laser display combining the PIC with a liquid-crystal-on-silicon (LCoS) panel, achieving 211% of the color gamut and more than 80% volume reduction compared to traditional LCoS displays. We further showcase its application in a see-through AR system. Our work represents a major advancement in the integration of nanophotonics with display technology, enabling a range of new display concepts, from high-performance immersive displays to slim-panel 3D holography.

# Main

In the history of display technology, the shift from bulky cathode ray tube (CRT) displays to compact flat-panel displays marked a pivotal moment. For nearly half a century, CRTs dominated the market until the advent of LED-based flat panel displays enabled a wave of portable devices, reshaping how we interact with visual technology in everyday life — from televisions to smartphones. Today, flat-panel displays are ubiquitous, yet the next leap forward — integrating laser technology into flat panels for higher brightness and richer color — remains a significant challenge.

Lasers offer superior brightness and color performance compared to conventional LED-based displays[1]. The high directionality of laser light affords projection of images with high brightness, which also enables lower-persistence displays critical for moving content. Their narrow spectrum provides more saturated colors, resulting in wider color gamut. Their polarized output reduces losses in display systems with polarization-sensitive components, leading to improved efficiency. These characteristics are particularly valuable for immersive experiences, such as augmented reality (AR), virtual reality (VR), and other high-performance display systems.

However, despite these benefits, current laser displays are largely confined to bulky projector formats, such as those used in movie theaters[2], where complex optical systems are used to deliver laser light to a screen. As shown in Figure 1a, a typical laser projector consists of multiple optical elements for beam expansion, beam shaping, color mixing, polarization control and other functions. Moreover, a significant volume for free space propagation is required to expand narrow laser beams over a large display area. Several attempts have been made to develop flat panel laser displays, but they require complicated laser arrays[3] or low-throughput fabrication methods[4], significantly limiting their performance and scalability.

The challenge of transitioning laser projectors to flat panels lies in the precise control required to manage laser light. Most LED-based flat panel displays, for example, liquid crystal displays (LCDs)[5], employ a 'diffuse and filter' approach, where light is first homogenized via random scattering in a thin lightguide, then passes through multiple layers of filters, such as pixelated color filters, polarizers, and angular filters to achieve the desired light output (Fig. 1c). This approach, however, cannot readily be applied to laser displays without negating the benefits of high directionality, polarization and color purity. Instead, laser displays require precise photon delivery, directing light exactly where it is needed with the right spatial, angular, spectral and polarization characteristics, to maintain their performance advantages. Achieving this level of precision necessitates complex bulky optical systems when using conventional optics, resulting in large projector systems.

Here, we introduce a new laser display architecture which features a slim, flat-panel form factor (Fig. 1b), overcoming the long-standing trade-off between size and performance in laser displays by using a large-scale visible photonic integrated circuit (PIC)[6-9]. We integrate thousands of components with different optical functions on a single photonic chip, carefully designed to cohesively control the red, green and blue (R,G,B) color channels to meet the high demands of advanced display applications. The PIC functions similarly to the lightguide in LCDs – expanding light and tailoring its spatial, angular, spectral and polarization characteristics according to the display system requirements. However, the working principle is very different: it adopts a 'guide and select' approach rather than the 'diffuse and filter' method. Light expansion is achieved via on-chip guiding and splitting, not random diffusing; angular spread is controlled by engineered grating emitters, not beam shaping filters; polarization and color purity are maintained and the light is out coupled selectively so that no additional color filters or polarizers are needed on the illumination side (Fig. 1d). Our architecture also offers new capabilities, such as on-demand chief ray angle (CRA) adjustment across the display, enabling co-design of the illumination module and optical system for maximum overall display efficiency. Fabrication of the PIC devices is based on standard CMOS-compatible processes, making it scalable for mass production. Laser-to-PIC integration[10,11], already established in the telecommunication industry, can also be adapted for the visible regime.

Our PIC illumination architecture provides a versatile platform for existing and emerging display technologies. It is compatible with a variety of display panels, including liquid crystal displays (LCD)[5], liquid-crystal-on-Silicon displays (LCoS)[12], digital light processing displays (DLP)[13], and has a wide range of applications, from smartphone displays to AR glasses[14] to emerging holographic displays[15-21]. As a proof of concept, here we demonstrate a flat panel laser display based on PIC and LCoS for AR applications, where compactness, high efficiency and high brightness are crucial.

**Display architecture**

The proposed flat panel laser display utilizes a non-emissive display configuration. Flat panel displays generally fall into two categories[22]: emissive displays, like OLEDs, which generate light directly from each pixel; and non-emissive displays, such as LCDs, where the pixels rely on external illumination rather than emitting light themselves. Here we choose a non-emissive configuration to avoid the complexity of integrating and controlling a large number of laser diodes on a dense, micron-scale pixel lattice. There are three key modules in non-emissive displays: the light source, the illumination unit, and the light modulation device. Here we use red, green and blue (R, G, B) semiconductor laser diodes; a PIC illuminator; and a color-sequential LCoS display panel respectively (Fig. 2a) to demonstrate this concept and its application for AR glasses.

The display stack is illustrated in Fig. 2a. The PIC device is placed over the LCoS cover glass, with a polarizer film laminated on top of the PIC device for image formation. LCoS is a reflective display which modulates the polarization upon reflection by changing the liquid crystal orientation at each pixel[12] (Fig. 2b). Through polarization filtering, the polarizer translates modulated polarization into modulated intensity, forming an image. To maximize the display resolution, we use a color-sequential LCoS without R, G, B subpixels – the full color image is created by showing the R, G, B color channels in rapid succession over a frame time. For quick prototyping, off-the-shelf fiber-coupled lasers are used, combined off-chip, and coupled to the PIC device through a single fiber attachment. Direct edge-coupling or flip-chip bonding of the lasers to the PIC and on-chip color mixing is discussed in Supplementary Section 3.

The display operation starts from light generation at the R, G, B laser diodes, which are turned on sequentially in synchronization with the LCoS color frame transition. Each laser is operated with a specific pulse width and duty cycle to optimize the spectrum and laser wall plug efficiency (WPE). To minimize the device footprint, we first combine the RGB colors, either on-chip or off-chip, and expand them using a shared set of cascaded Y-splitters (Fig. 2c). Following the beam expansion, the white light is separated back into R, G, B colors via coarse-wavelength-multiplexers (CWMs) of the same design but in the reverse direction. Here color splitting is needed for individual control of the R, G, and B emission profiles. In the emission region, light is extracted via arrays of pixelated grating couplers (Fig. 2d). Each grating emitter is designed to be 1~2 μm in length to produce a finite diffraction cone angle, which can be matched with the numerical aperture (NA) of the viewing optics (Fig. 2d, 2e). The emission area is slightly oversized compared to the active area of the LCoS panel to avoid brightness roll-off at the edges. Finally, the illumination light is modulated by the LCoS panel upon reflection, and forms an image after the polarizer. Figure 2f-i shows the optical images of the wafer, the SEM images of the PIC stack and Y-splitters, respectively. More details on the PIC fabrication process can be found in Methods and Supplementary Section 1.

**Design principle**

Our design carefully manages the trade-offs in the four key performance metrics for the PIC illuminator: uniformity, polarization extinction ratio, transparency and efficiency, which correspond directly to the critical display performance factors: display uniformity, contrast ratio, ghosting, and power consumption, respectively (Fig. 3a). Here uniformity includes both brightness and color.

One significant trade-off is between brightness uniformity and light extraction efficiency: stronger gratings increase extraction efficiency but degrade uniformity due to exponential intensity decay along the waveguides. To overcome this, we introduce a novel optical circuitry design — spatial interleaving of waveguide circuits. As illustrated in Fig. 2c, light is split on-chip and fed from opposite directions via two sets of Y-splitters. This creates a compensatory effect, maintaining uniform illumination despite the monotonic decay in each direction (Fig. 3a).

We achieve high color uniformity by optimizing the PIC layer stack and grating strength for each color. According to the Born approximation, the grating outcoupling efficiency is proportional to the mode intensity at the grating structures. Therefore, to realize good color uniformity requires matching the grating efficiency for R, G, and B, which in turn requires engineering their mode profiles. To achieve this, we use a three-layer PIC design – the gratings are patterned in a separate aluminum oxide ($AlO_x$) layer above the silicon nitride (SiN) core waveguiding layer, with a silicon dioxide ($SiO_2$) spacer layer in between (Fig. 2d, 2e). Compared to the conventional direct waveguide etching, the three-layer stack offers more design degrees of freedom to tailor the mode and grating interaction as a function of wavelength.

The trade-off between uniformity and efficiency is shown in Fig. 3b-c, with a specific focus on the spacer layer thickness $t_s$ for illustration purposes. Following display industry convention, brightness uniformity is defined by the min/max brightness ratio, which must exceed 80%. Color uniformity, measured by $\Delta u'v'$ (deviation from the white point)[23], must remain below 0.01. One can clearly see the design trade-offs: the efficiency and brightness uniformity trend oppositely with $t_s$, while color uniformity is optimized at $t_s$ around 200 nm. In our device, we choose $t_s$ = 170 nm as it offers the overall best performance: it can meet the uniformity requirements while achieving a high light extraction efficiency of 60%. More detail on the further potential trade-off optimization on the uniformity is discussed in Supplementary Section 3 and 4.

Polarization extinction ratio (PER), critical for display contrast, is optimized by using single-mode waveguides and polarization-selective gratings. As shown in Fig. 3a, the illumination polarization is designed to be orthogonal to the polarizer transmission axis. Any cross polarization will lead to dark state light leakage and thus low display contrast. Here we achieve an on-axis PER above 10,000:1 in simulation (Fig. 3i) by carefully ensuring high polarization selectivity in the waveguides and grating couplers. Note that the PER decreases at oblique angles, creating a trade-off between contrast and the numerical aperture (NA) of the optical system (Fig. 3f-3h).

Due to the reflective nature of the LCoS, light passes through the PIC illuminator again upon reflection from the LCoS panel. This adds another key requirement – the PIC must be highly transparent to avoid any ghosting or image quality degradation. Note that the PIC's transparency in the imaging path and high efficiency in the illumination path do not violate reciprocity since the emitted light is diverging. Upon reflection, the beam size has grown significantly beyond the effective grating cross-section. In fact, the re-interaction of light with PIC is dominated by diffraction at the ridge waveguides, rather than at the gratings. Figure 3j shows the simulated diffraction efficiency integrated over the collection cone angle. Since both SiN and $AlO_x$ layers are optically thin, the diffraction efficiency from the PIC waveguides is less than 0.8% across the spectrum of interest, meeting the system ghost specification of <1%.

Finally, we optimize the optical efficiency of the PIC device by refining all components, minimizing propagation and bending losses in waveguides, insertion losses at Y-splitters and CWMs, optimizing laser-to-PIC coupling efficiencies, designing impedance mismatches at grating emitters, and maximizing grating extraction efficiency (see Supplementary Sections 2-4). A low-loss SiN platform has been previously demonstrated and reported across visible to near-IR wavelength regimes[24-28]. In this work, we push these boundaries further, achieving record-low propagation losses for R, G, B wavelengths (635, 520, 450 nm) down to 0.1 dB/cm, 0.3 dB/cm, and 1.1 dB/cm, respectively. The SiN core thickness is carefully designed to minimize bending and insertion losses while ensuring single-mode operation across all colors and maintaining compactness. The Y-splitters are inversely designed for minimal insertion loss and broadband performance. With a series of comprehensive designs and further process optimizations, we anticipate our architecture could achieve more than 10-fold improvement over conventional LCoS projectors (see Supplementary Section 5).

**Results**

We have evaluated the performance of our flat-panel laser display architecture in a series of experiments, starting with the characterization of the stand-alone PIC illuminator, followed by the assembled flat-panel laser display, and finally a full AR system.

High illumination uniformity is achieved experimentally. Figure 3d shows the measured illumination field before integrating with the LCoS. The PIC emission area is 6mm by 4.8mm, with the long axis along the waveguides. Due to the interleaved light propagation, the illumination is brighter towards the edges and dimmer towards the center. The measured brightness uniformity (min/max) is 71%, slightly lower than the design value, due to the layer thickness variation from fabrication shift, meanwhile a good color uniformity with $\Delta u'v'<0.01$ is achieved. Note that the human perception of color variation depends strongly on the spatial frequency, i.e., slowly varying color change across a large field of view (FoV) is much less noticeable. Therefore, while the color variation is quite visible in the figure, it is perceptually less obvious when observed over the designed 30-by-40 degree FOV. A zoomed-in view of the PIC is shown in Figure 3e, showing the individual grating emitters. Note that due to circuitry interleaving, the emitter lattice spacing in the vertical direction (along waveguides) is twice that in the horizontal (transverse to waveguides) in this design.

We have also measured the illumination polarization extinction ratio (PER) (Fig. 3i). Due to limitations in the characterization setup, the measured PER is integrated over a numerical aperture (NA) of 0.2. A spatial variation of the measured PER across the PIC device is observed, mainly due to the unwanted light scattering at the edge coupling region (Supplementary Section 4). Away from the edge coupler, an illumination contrast of more than 250:1 at NA~0.2, in agreement with the simulation.

Next, we evaluated the display performance after the PIC-LCoS integration. Figure 4a shows a fully functional display assembly with a compact form factor and high brightness. The LCoS panel used here has a 4.5 μm pixel pitch with FHD (1920×1080) resolution and operates at 120 Hz video frame rate.

One advantage of laser displays is wider color gamut. As we use narrow band prime color lasers, the color gamut is wider than any other LED-based displays (Fig. 4b). It provides 211% Gamut ratio and full coverage of sRGB/BT. 709[29], the standard for high-definition TV defined by the International Telecommunication Union (ITU). As a comparison, our display encompasses 74% of the visible colors specified by the CIELAB color space, while the standard Adobe RGB color space[30] covers just 52% and sRGB[31] covers only 36%. Note that the current device does not fully cover Adobe RGB because the green primary color (center wavelength at 515 nm) is different from that in LED displays (center wavelength around 530-540 nm) due to the availability of green laser diodes. In the future, green laser diodes with longer center wavelengths can be used to further improve the color performance.

The display performance under direct view is shown in Fig. 4c-4f. An eyepiece is used to magnify the displayed image (see Methods). The image blur towards the edge is due to the aberrations of the eyepiece lens, and not from the display. The assembled display sequential contrast is around 40:1, significantly lower than the illumination PER. This is limited both by the LCoS panel contrast, and the alignment error between the PIC, LCoS and the polarizer film which was done manually for this demonstration. In actual manufacturing, this issue can be mitigated by a more precise alignment process based on optimized alignment marks.

A compact and efficient display engine is particularly important for augmented reality (AR) glasses, given their tight space and power budget limited by the glass size and weight. Here we demonstrate the PIC-laser display performance at a system level by pairing it with an off-the-shelf pupil-replicating AR geometrical glass (Fig. 4g). Light from the flat panel laser display is first collimated by a custom-made lens module (Supplementary Section 6), then coupled to the input pupil of the AR lightguide. The exit pupil is expanded through multiple total internal reflections (TIRs) inside the glass plate (Supplementary Section 9). The current demonstration is a handheld AR setup with the projector supporting 50 deg diagonal FoV. The volume mainly comes from the large off-the-shelf LCoS display driving board, and can be miniaturized significantly with a customized ASIC-based driver. With better integration and packaging, the light engine size can be reduced to under 1 cubic centimeter (Supplementary Section 6), enabling ultra-compact and lightweight AR experience (Fig. 4h). Several types of AR use cases are demonstrated. Figure 4i demonstrates the mixed reality experience where the virtual objects are blended seamlessly with the real world scene in an office environment. Instant messaging and notifications are illustrated in Fig. 4j-k respectively. Note that the speckles are much less noticeable after the AR lightguide, partially due to the reduced additive contrast with the see-through background.

**Discussion**

Realizing the full potential of PIC laser displays still faces several challenges. First is the laser speckle. Speckles are a known issue in laser displays[1]. Here we observe them as granular patterns in the image, resulting from the interference of different PIC emitters on the LCoS plane. As a first proof-of-principle demonstration, no additional despeckling method is implemented in the current device. We measured around 20% speckle contrast with a 1 nm bandwidth single-mode laser, exceeding the typical perception threshold of 4%[32]. Multiple mitigation methods such as wavelength or polarization diversity[33], dynamic diffusers[34] and microlens arrays[35], have been proposed and tested in existing laser projectors, and can be adapted to the PIC laser displays. Results on despeckling are being prepared for publication in a separate manuscript.

Another key challenge is light source integration and packaging. While laser integration is mature in silicon photonics for telecommunications wavelengths[10,11], the integration of RGB laser diodes with visible photonics is still at an early stage[27,36]. We have demonstrated direct laser-PIC edge coupling at a prototype level via active alignment (Supplementary Section 10). For mass production, die-to-wafer flip-chip bonding can be used as a cost-effective scalable manufacturing path, and recent progress on heterogenous visible laser integration with SiN photonics provided another possibility on this avenue[37].

For AR application specifically, there is a challenge of the power waste and contrast degradation for sparse content in many use cases such as depicted in Fig. 4i-k. Currently the PIC illuminator can only be turned on and off globally, leading to unfavorable power consumption when only a small portion of the LCoS display is on[38,39]. To further improve the illumination efficiency, active PIC modulation[6,40-44] might be used to control different illumination regions independently (Fig. 5a), a concept similar to local dimming in LCDs[45].

Beyond AR/VR, the PIC illuminator serves as an enabling platform for a broad range of new display concepts, including slim-panel holographic displays[15,16], high-resolution light field displays[46], pupil-steered displays[47], and many more. As an example, Fig. 5b illustrates a potential application of PIC in holographic displays, where the PIC illuminator is integrated with spatial light modulators (SLMs) and a holographic pancake lens[48]. Unlike previous holographic displays which rely on simple plane wave incidence[18,49], the PIC could provide a tailored illumination field co-optimized with the holographic display system, achieving unprecedented compactness, brightness and immersiveness.

This work introduces large-area visible photonic integrated circuits (PICs) as a powerful platform for compact and efficient laser illumination. We demonstrate a flat-panel laser display architecture that achieves substantial miniaturization, paving the way for advanced applications in augmented reality and beyond. We believe our work could facilitate more innovations that leverage advanced nanophotonics within the display industry, driving advancements in next-generation visual technologies.

**Method**

*Device fabrication and integration*

The photonic integrated circuit (PIC) device is fabricated in a 200 mm (8 inch) CMOS foundry. The fabrication process begins with 200 mm silicon (Si) wafers coated with a 2.5μm-thick thermal oxide (SiO2) layer, followed by the deposition and patterning of a low-pressure chemical vapor deposition (LPCVD) silicon nitride (SiN) layer, serving as the core waveguide. Deep-ultraviolet (DUV) lithography and reactive ion etching (RIE) are employed for this patterning. Subsequently, a spacer $SiO_2$ layer is deposited over the SiN and planarized using chemical-mechanical polishing (CMP). Aluminum oxide ($AlO_x$) is then deposited via atomic layer deposition (ALD) and patterned to form grating emitters. Another $SiO_2$ layer is deposited and planarized thereafter. Titanium nitride (TiN) is deposited and patterned to function as a scattered and stray light absorber. Finally, a top layer of $SiO_2$ is deposited over the TiN. The device layers are transferred to a quartz substrate through wafer bonding, followed by the removal of the Si handle wafer. Additional details on the fabrication process can be found in Supplementary Section 1.

The fabricated wafer is singulated into individual dies, which are then edge-polished for fiber or direct laser coupling. The fiber-attached device chip is aligned and bonded onto a commercial LCoS panel using UV-curable epoxy. A thin-film reflective polarizer with pressure-sensitive adhesive (PSA) is affixed to the top side of the PIC chip, serving as a polarization analyzer. For augmented reality (AR) demonstration, the packaged PIC-LCoS sample is further aligned with a custom-designed projector lens and a geometrical waveguide combiner. (See Supplementary Sections 7-9)

**Author contributions:**

Zhujun Shi, Risheng Cheng, Guohua Wei worked on the architecture, design, simulation, planning and analyzed the results; Min Chul Shin worked on the design and simulation of the device; Steven Hickman worked on the device fabrication; Peter Topalian worked on the device integration and prototyping; Lei Wang worked on the laser sources; Dusan Coso worked on the laser integration. Alexander Koshelev, Maxwell Parsons contributed to the ideation and initial demonstration of the concept; Brian Le, Lizzy Lee worked on the device characterization; Sean Braxton worked on the projector design; Rahul Agarwal contributed to the initial demonstration; Barry Silverstein, Yun Wang and Giuseppe Calafiore supervised the project.

**Acknowledgments**

The authors would like to thank Zhimin Shi, Jinran Qie, Mohan Shen, Joshua Zhao, Arnab Manna for the various discussions around PIC architecture and component design, Michael Schaub for projector optical design studies, Lindsey Gilman for thermal analysis, Kailing Zhang, Andrew Copsey for PIC device fabrication process development, Luan Nguyen, Yuhang Wu for data collection, Simone Carpenter, Andrew Gunderson, Michael Beltran, for optomechanical support, John DeFranco for project management support. Zhujun Shi would like to thank Dr. Zhehao Dai. We appreciate our partner TDK for their work on laser/PIC integration.

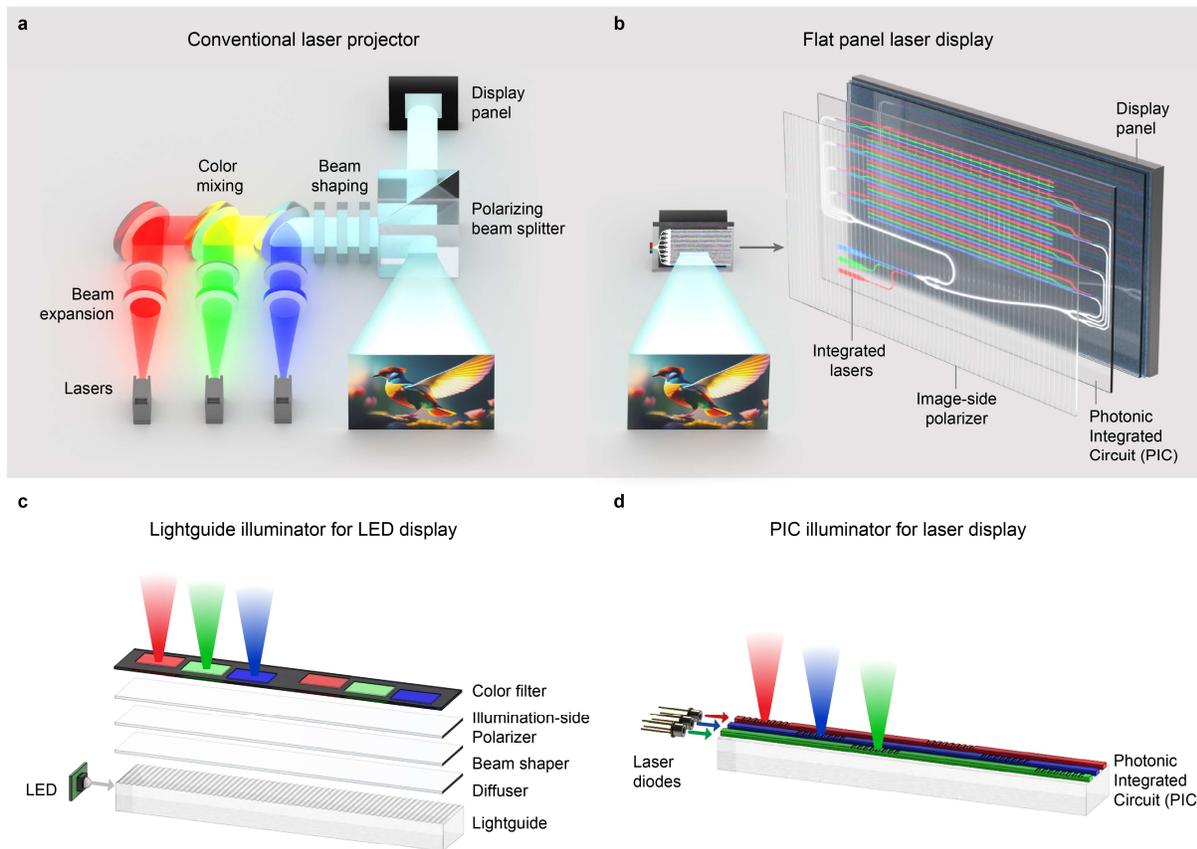

Figure 1. Concept of flat panel laser displays. (a) Schematics of a conventional laser projector using free-space illumination. It consists of laser sources, collimating lenses, dichroic mirrors, beam shaping elements, a polarizing beam splitter and a display panel. (b) Schematics of the proposed flat panel laser display. A photonic integrated circuit (PIC) is used to replace the free space illumination module and integrated directly onto the display panel, achieving a compact flat panel form factor. (c-d) Comparison of LED and laser illumination. (c) Typical LED lightguide illuminators, or backlights, use multiple layers of diffusers and light filters to tailor the spatial, angular, spectral and polarization characteristics of light, resulting in low optical efficiency. (d) The PIC illuminator eliminates the need for lossy diffusers and filters by guiding and tailoring the light characteristics on-chip.

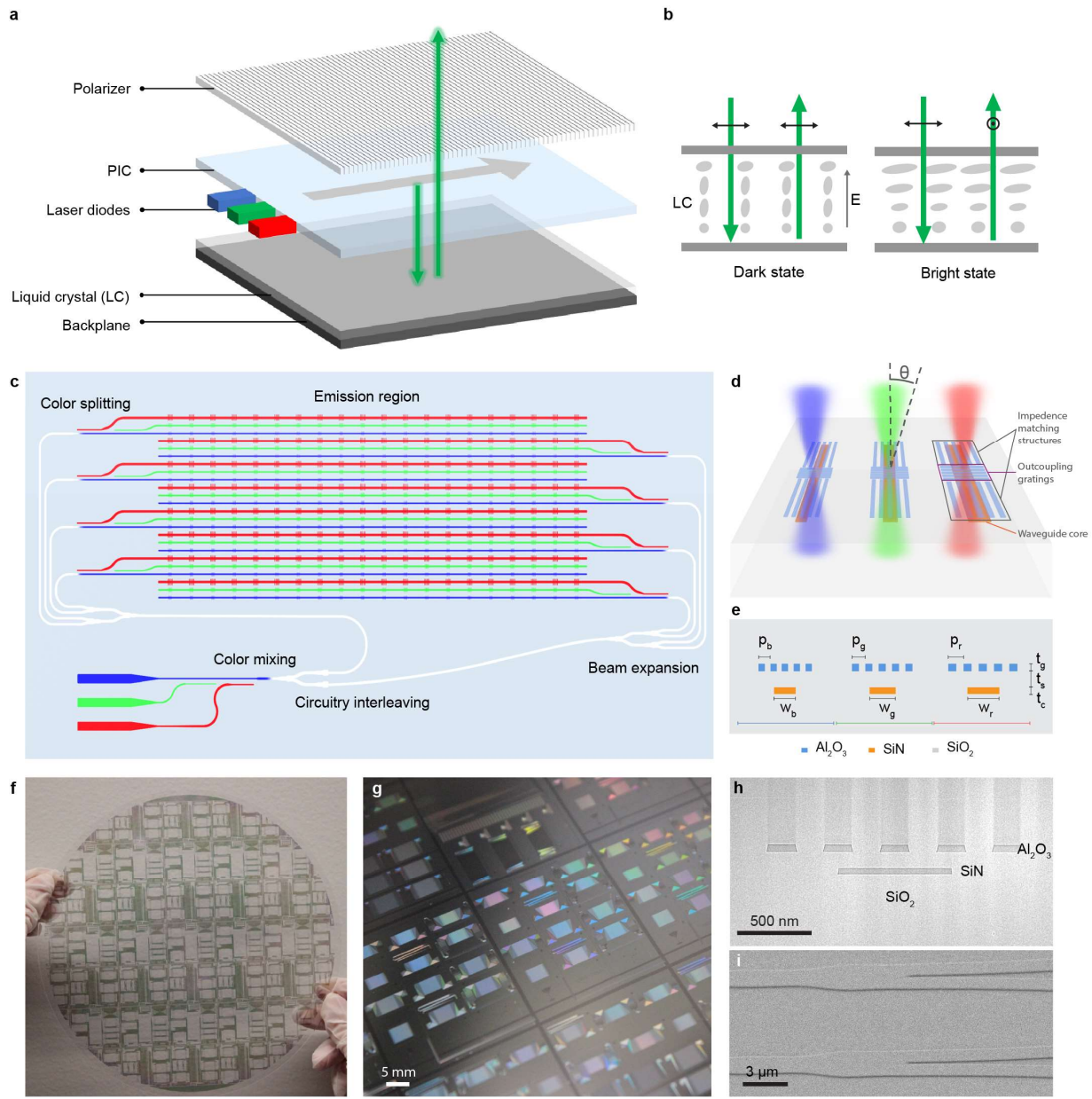

Figure 2. (a) Schematics of the flat panel laser display stack. The PIC is placed between a liquid crystal on Silicon (LCoS) display panel and a polarizer. It expands the laser inputs over an area to illuminate the LCoS panel. The LCoS then reflects the light and modulates its polarization spatially, which is converted into intensity modulation after passing through the polarizer. (b) Polarization rotation at the liquid crystal (LC) layer. In the dark state, when the voltage is applied, the LC molecules are aligned vertically and do not significantly alter the polarization of the incoming light. As a result, most of the reflected light is blocked by the cross-polarizer, creating a dark display area. In the bright state, when no voltage is applied, the liquid crystal molecules

naturally form a twisted structure that rotates the polarization of the incident light by 90 degrees upon reflection. This allows the light to pass through the polarizer, resulting in a bright state. (c) Schematics of the PIC layout. (d) Zoom-in view of the grating emitters. The R,G,B emitters are placed side-by-side, sharing the same core and grating layers. For each emitter, the grooves transverse to the core waveguide are short pixelated gratings for light extraction. The grating pitch determines the emission direction, the grating length determines the divergence angle. The grooves parallel to the core waveguide are impedance matching structures. They are introduced between adjacent emitters to minimize the scattering loss. (e) Cross-section view of the PIC stack. It consists of a 50 nm SiN core, a 170 nm $SiO_2$ spacer, and a 55 nm $AlO_x$ grating layer. The core waveguide widths and the grating pitches are optimized separately for R,G,B. (f) Optical image of a 200 mm diameter PIC wafer. (g) Optical image of the device before the wafer is transferred to the glass substrate (see Method). (h) SEM image of the cross-sectional layer stack. (i) SEM image of the Y-splitters.

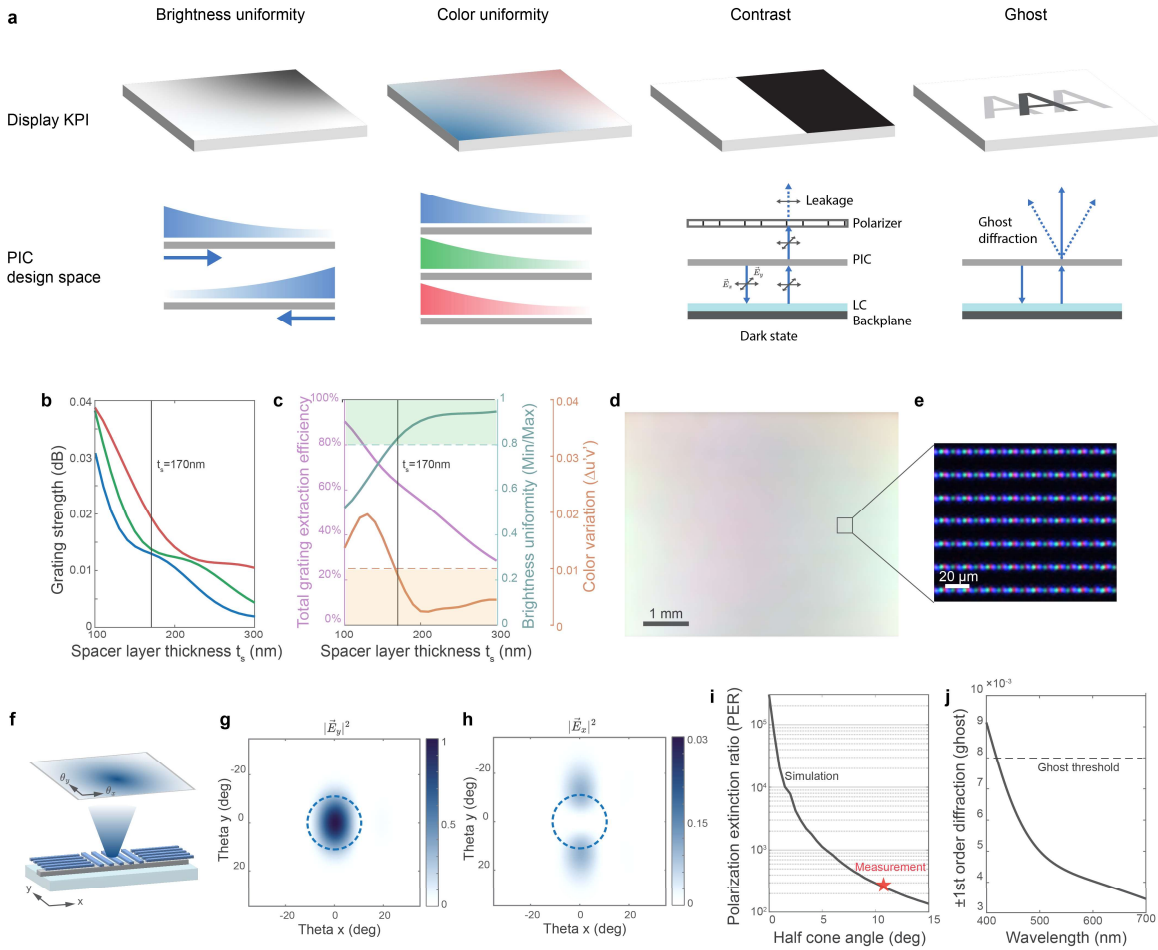

Figure 3. (a) Key display performance factors and the corresponding PIC design space. (b) Simulated grating strength as a function of the SiO$_2$ spacer layer thickness $t_s$ for R,G,B. The SiN core thickness and the AlO$_x$ grating layer thickness are fixed at 50 nm and 55 nm respectively. (c) Simulated brightness uniformity, color uniformity and light extraction efficiency as a function of $t_s$. The green shadow represents the allowable range for brightness uniformity, Min/Max>0.8. The orange shadow represents the allowable range for color uniformity, $\Delta u'v'$<0.01. (d) Measured uniformity map of a fabricated PIC illuminator. The device size is 6 mm by 4 mm. (e) Zoom-in view of the PIC emitters. (f) Schematics of the grating emitters. (g) Simulated far field intensity profiles for the Y-polarization (transverse to the polarizer transmission axis). (h) Simulated far field intensity profiles for the X-polarization (parallel to the polarizer transmission axis). (i) Gray line: simulated polarization extinction ratio (PER) as a function of the light collection half angle. Red star: measured PER at 11 deg half cone angle. (j) Simulated diffraction efficiency into the ghost orders.

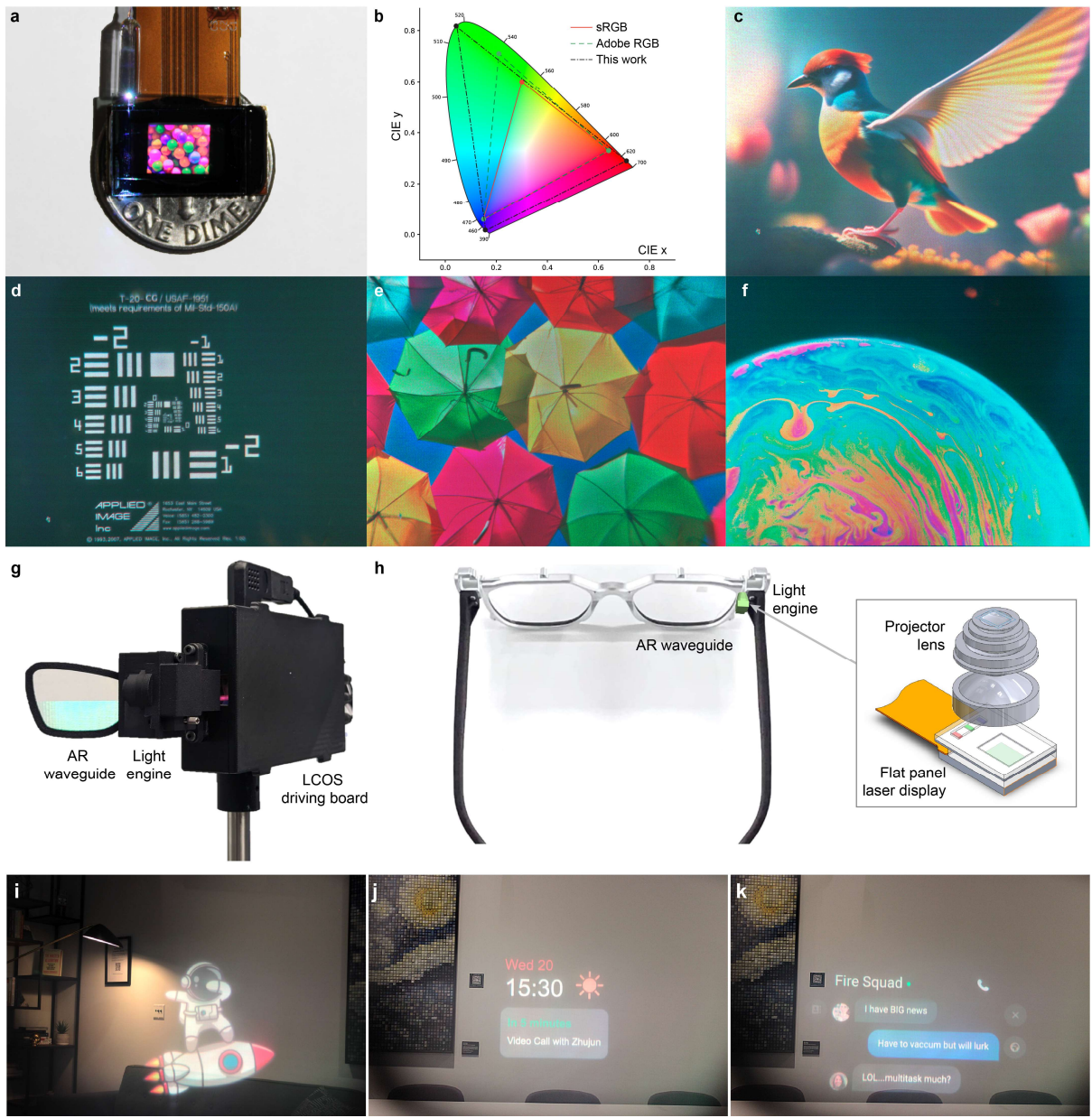

Figure 4. (a) The assembled flat panel laser display shown on top of a dime coin. A fiber is attached to the PIC for light input. The flex cable is for LCoS control. The light sources and laser controllers are not shown in the photo. (b) Color performance of this work in comparison to the standard color space. (c-f) Measured display images. (g) Photo of the handheld AR setup used in the experiment. (h) Conceptual visualization of an AR system integrated with a form-factor optimized flat panel laser display. (i-k) Images captured by a camera at the eye position of the AR setup in Fig. 4g, showing the displayed virtual objects and real world of an office.

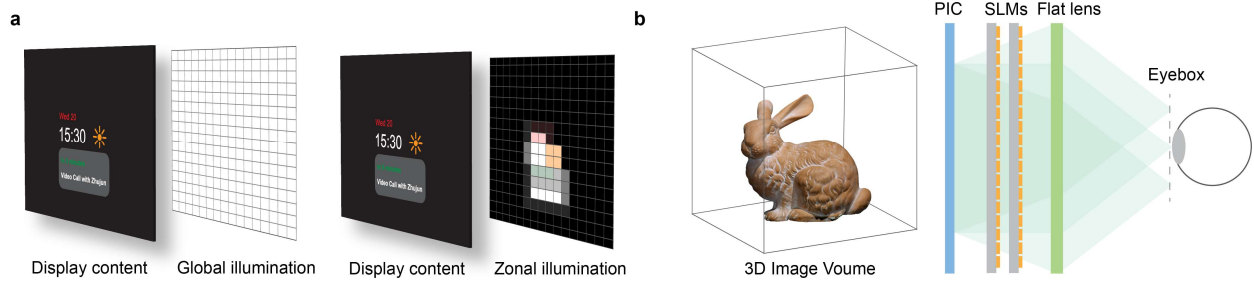

Figure 5. (a) Concept of zonal illumination. Zonal illumination can enhance image contrast and efficiency by turning on the illumination only where needed. It can be realized by using active PIC modulators. (b) Schematics of a PIC-enabled holographic display. The PIC provides a tailored illumination field, which feeds into one or multiple layers of spatial light modulators (SLMs) for hologram generation. A flat lens, such as a holographic optical element, is then used to project and magnify the image.